\documentclass[11pt,twoside]{article}

\usepackage{asp2006}
\usepackage{epsf}
\usepackage{psfig}
\usepackage{lscape}

\begin{document}
\title{Stellar Velocity Profiles and Line-Strengths out to Four Effective Radii in the Early-Type Galaxy NGC\,3379}  
\author{A. Weijmans\altaffilmark{1}, M. Cappellari\altaffilmark{2}, P.T. de Zeeuw\altaffilmark{3,1},  E. Emsellem\altaffilmark{4}, \\J. Falc\'on-Barroso\altaffilmark{5}, H. Kuntschner\altaffilmark{3},  R.M. McDermid\altaffilmark{6},\\ R.C.E. van den Bosch\altaffilmark{7},  and G. van de Ven\altaffilmark{8}}  
\affil{\altaffilmark{1}{Sterrewacht Leiden, Leiden University, Postbus 9513, 2300 RA Leiden, the Netherlands [weijmans@strw.leidenuniv.nl]}\\
\altaffilmark{2}{Sub-Department of Astrophysics, University of Oxford, Denys Wilkinson Building, Keble Road, Oxford OX1 3RH, UK}\\
\altaffilmark{3}{ESO, Karl-Schwarzschild-Str 2, 85748 Garching, Germany}\\
\altaffilmark{4}{CRAL, University of Lyon, 9 Avenue Charles Andr\'e, 69230 Saint Genis Laval, France} \\
\altaffilmark{5}{ESTEC, Postbus 299, 2200 AG Noordwijk, the Netherlands}\\
\altaffilmark{6}{Gemini Observatory, Northern Operations Centre, 670 N. A'ohoku Place, Hilo, Hawaii 96720 USA}\\
\altaffilmark{7}{McDonald Observatory, University of Texas, Austin, TX 78712, USA}\\
\altaffilmark{8}{IAS, Einstein Drive, Princeton, NJ 08540, USA}
}

\begin{abstract} 
We describe a new technique to measure stellar kinematics and
line-strengths at large radii in nearby galaxies. Using the
integral-field spectrograph SAURON as a 'photon-collector', we obtain
spectra out to four effective radii ($R_e$) in the early-type galaxy
NGC\,3379. By fitting orbit-based models to the extracted stellar
velocity profile, we find that $\sim$ 40\% of the total mass within
5 $R_e$ is dark. The measured absorption line-strengths reveal a
radial gradient with constant slope out to 4 $R_e$.
\end{abstract}


\section{Introduction}

Although the presence of dark matter dominated haloes around spiral
galaxies is well established \citep[see e.g.][]{va85_aw}, little is
known about the dark haloes around early-type galaxies. Large regular
{\sc H\,i} discs or rings, whose kinematics are often used to constrain
the properties of dark haloes, are rare in these systems
\citep*[though see][]{fr94_aw,we08_aw}, so that we are forced to use
other tracers of the gravitational potential.

Here we use stellar kinematics obtained with the integral-field
spectrograph SAURON \citep{ba01_aw} at large radii in the elliptical
(E1) galaxy NGC\,3379 to model its dark halo. This galaxy is of
intermediate luminosity ($M_B = -20.57$) and has a half-light or
effective radius $R_e$ of 42 arcsec, which corresponds to 2.1 kpc at its
distance of 10.3 Mpc.

\section{Method}

\begin{figure}
\begin{tabular}{cc}
\psfig{figure=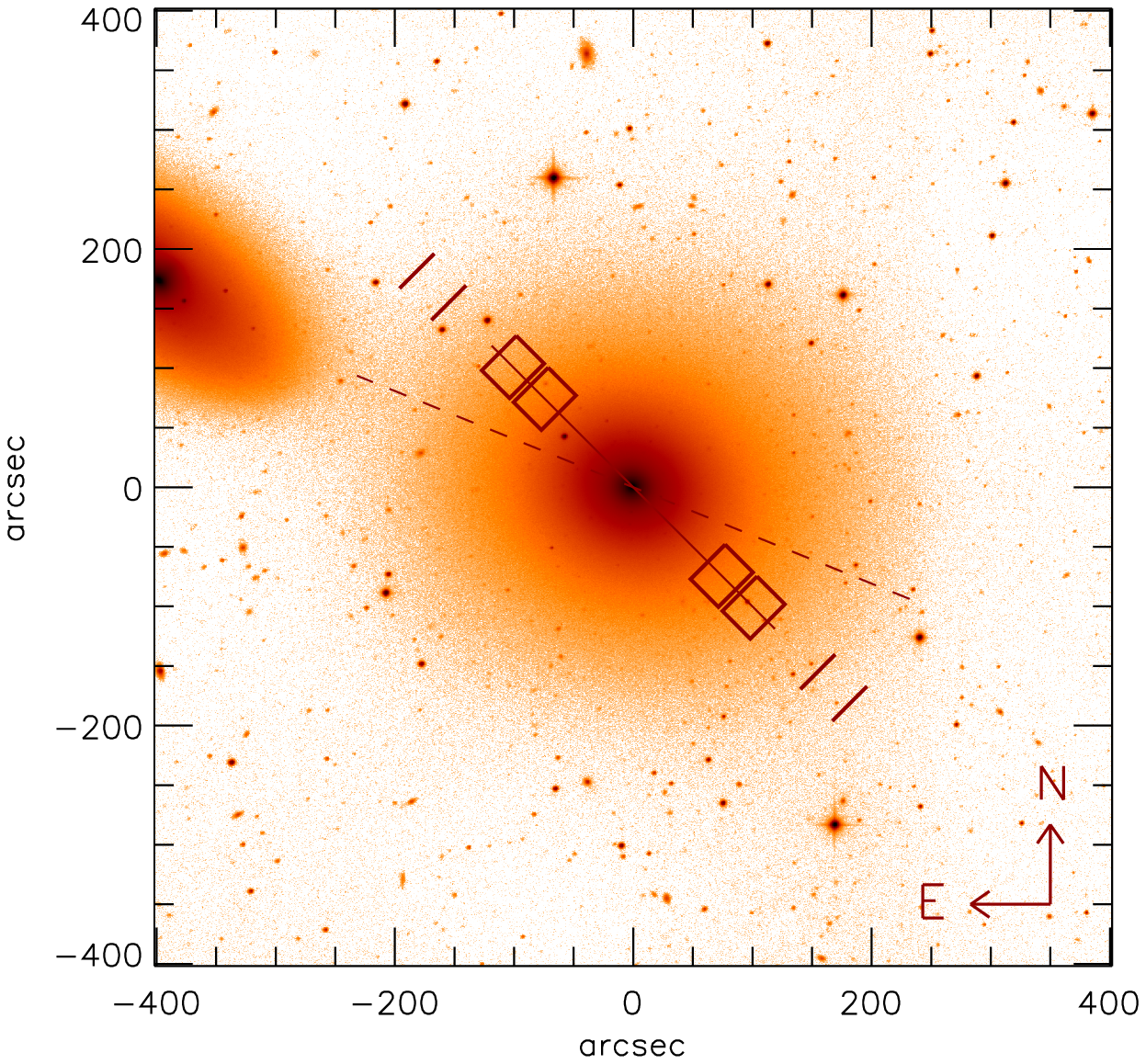,angle=0,width=5.0cm,clip=true}  & 
\psfig{figure=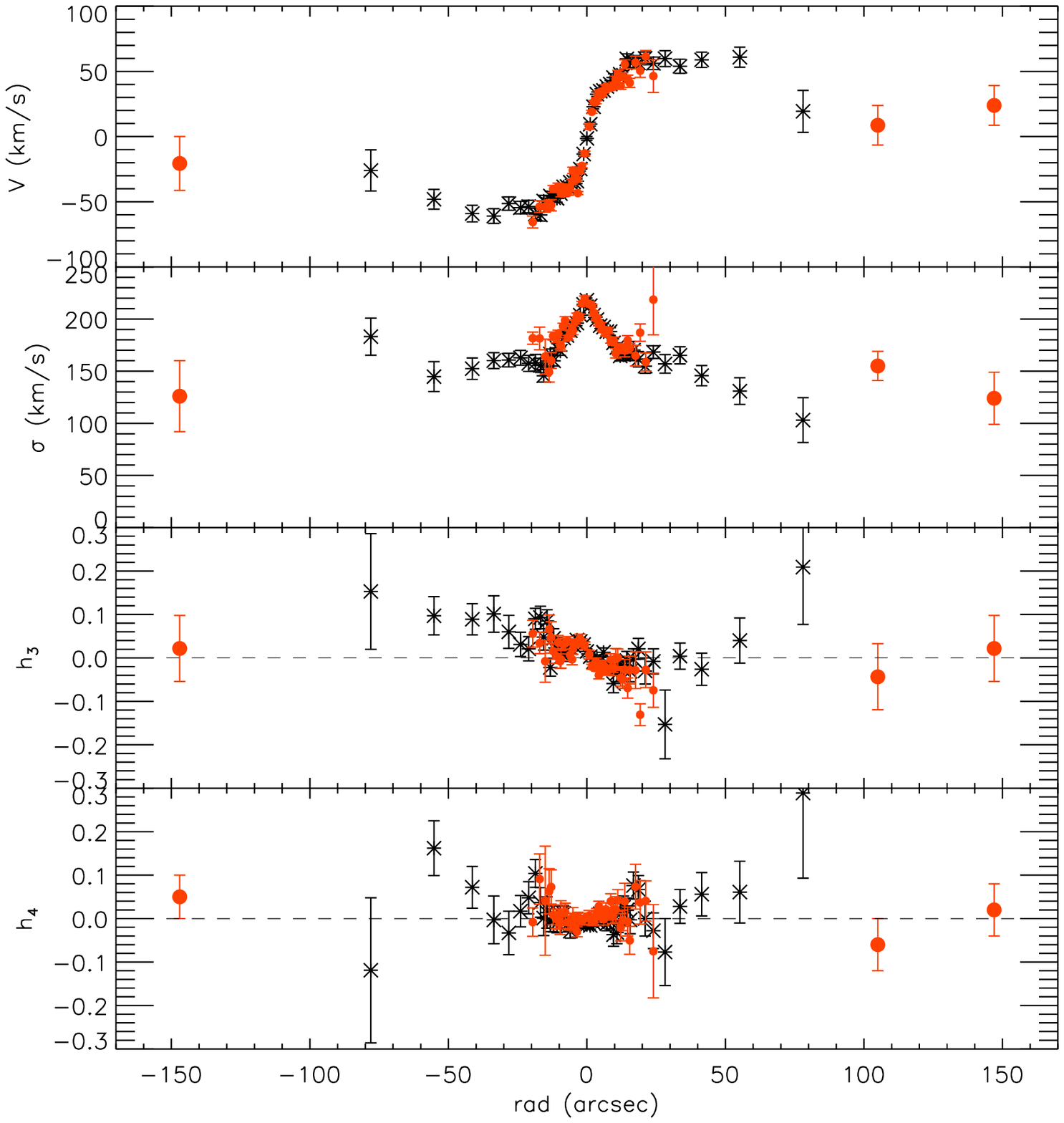,angle=0,width=7.0cm} \\
\end{tabular}
\caption{Left: Positions of our observed fields in NGC\,3379. The red
boxes denote each SAURON field-of-view. The skylenslets (red short
thick lines) are aligned with the long side of the SAURON field, at a
distance of two arcminutes. The dashed line denotes the major axis of
NGC\,3379. The underlying $V$-band image was obtained with the 1.3-m
McGraw-Hill Telescope at MDM Observatory. Right: LOSVD of NGC\,3379
out to 4 $R_e$. The black stars are long-slit data from
\citet{st99_aw} and the central red dots are SAURON observations
obtained in the original survey \citep{em04_aw}. The red dots at large
radii are our new observations, and double the spatial extend of the
data. }
\label{fig:pos_ngc3379}
\end{figure}

We centred SAURON at 2.6 and 3.5 $R_e$ on both sides of the nucleus of
NGC\,3379, close to its major axis (see Fig.~\ref{fig:pos_ngc3379},
left panel). A single spectrum of one lenslet is dominated by noise at
these large radii, but adding all spectra of all lenslets together we
obtained in three out of our four fields sufficient signal-to-noise to
measure the stellar absorption line-strengths and the line-of-sight
velocity distribution (LOSVD) up to the fourth Gauss-Hermite moment
$h_4$. This last parameter is necessary to break the well-known
mass-anisotropy degeneracy when modeling the dark halo
\citep[e.g.][]{ge93_aw}.

\section{Results}

We measured the LOSVD using the penalized pixel fitting method (pPXF)
by \citet{ca04_aw}. The resulting LOSVD (Fig.~\ref{fig:pos_ngc3379},
right panel) shows a smooth continuation of existing stelllar
kinematic measurements \citep{st99_aw}.

\begin{figure}
\centerline{\psfig{figure=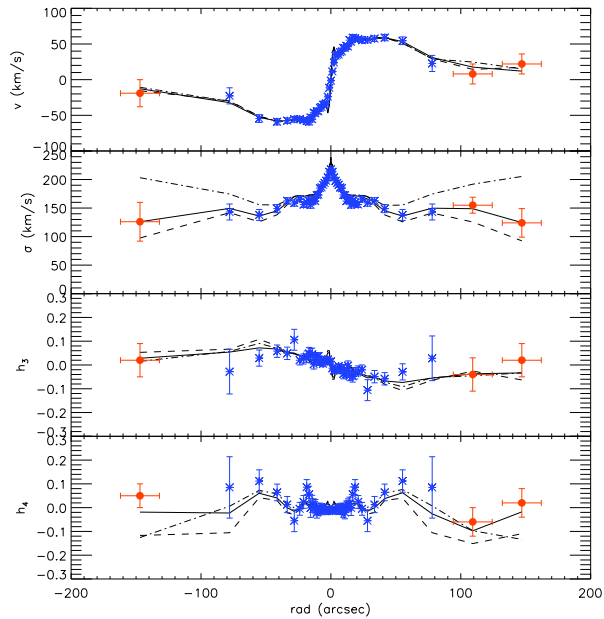,angle=0,width=8.0cm}} 
\caption{Best-fit model (solid line) overplotted on datapoints. The blue stars are (symmetrized) long-slit data from \citet{st99_aw} and the red dots are our datapoints (horizontal error bars indicate the width of the SAURON field-of-view). Also shown are a model without a dark halo (dashed line) and a model with a ten times too massive halo (dot-dashed line). These models do not fit the observed dispersion and $h_4$ profiles.}
\label{fig:halo}
\end{figure}

We use a Schwarzschild model \citep*{vb08_aw,vv08_aw} to fit our
measurements, including the central SAURON field of the original
survey \citep{em04_aw} and the long-slit data of \citet{st99_aw}. The
black hole mass and the (nearly axisymmetric) shape of the stellar
distribution of NGC\,3379 are taken from \citet{vbthesis_aw}. We model
the spherical dark halo with an NFW profile \citep*{na96_aw} with a
concentration $c=10$ \citep{bu01_aw}. Our best-fit model is shown in
Fig.~\ref{fig:halo}, and has a total halo mass $M_{200}$ of $1.0
\times 10^{12}$ $M_\odot$, which corresponds to 10 times the total
stellar mass of NGC\,3379.

\begin{figure}
\begin{tabular}{cc}
\psfig{figure=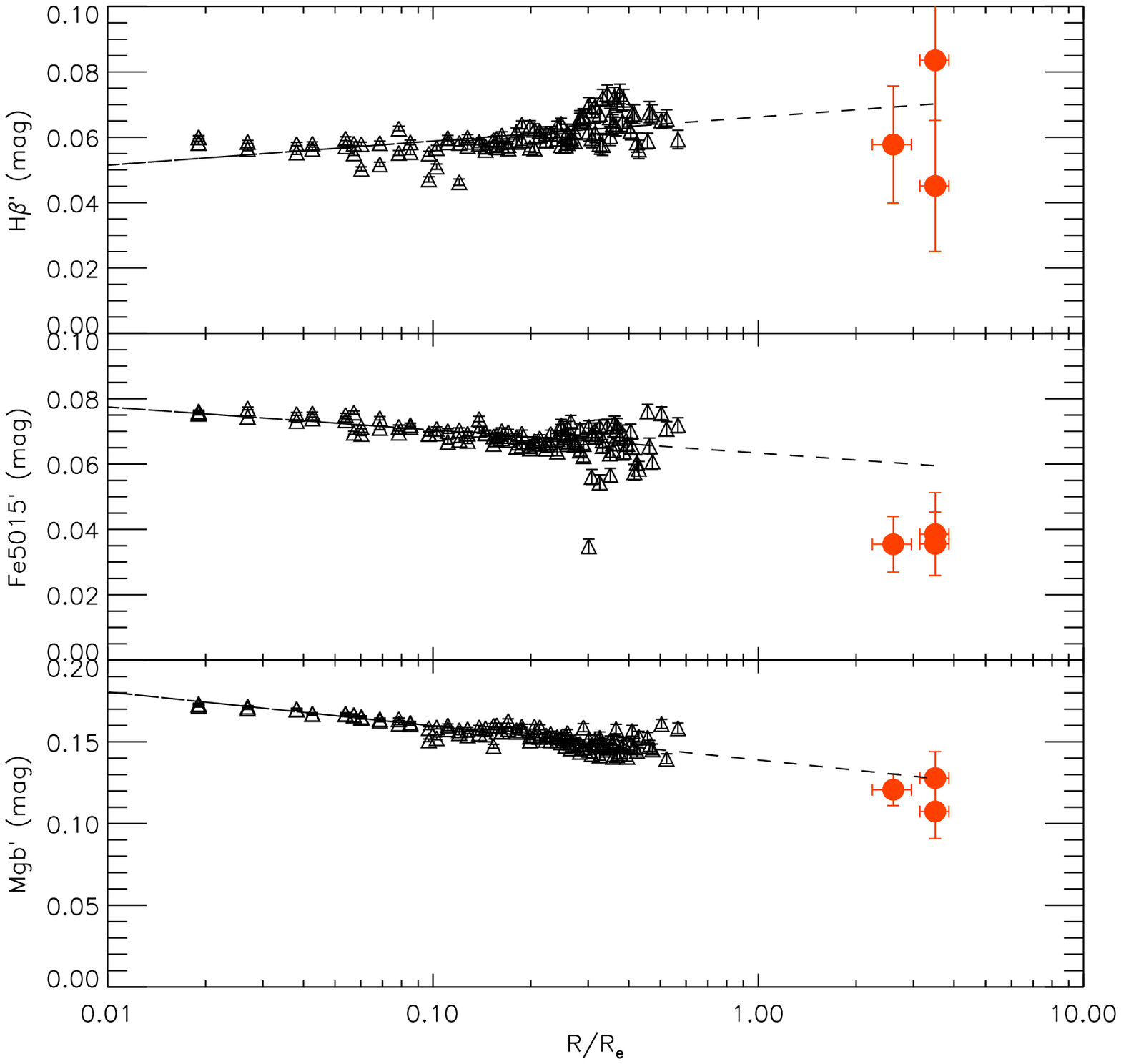,angle=0,width=5.3cm} & \psfig{figure=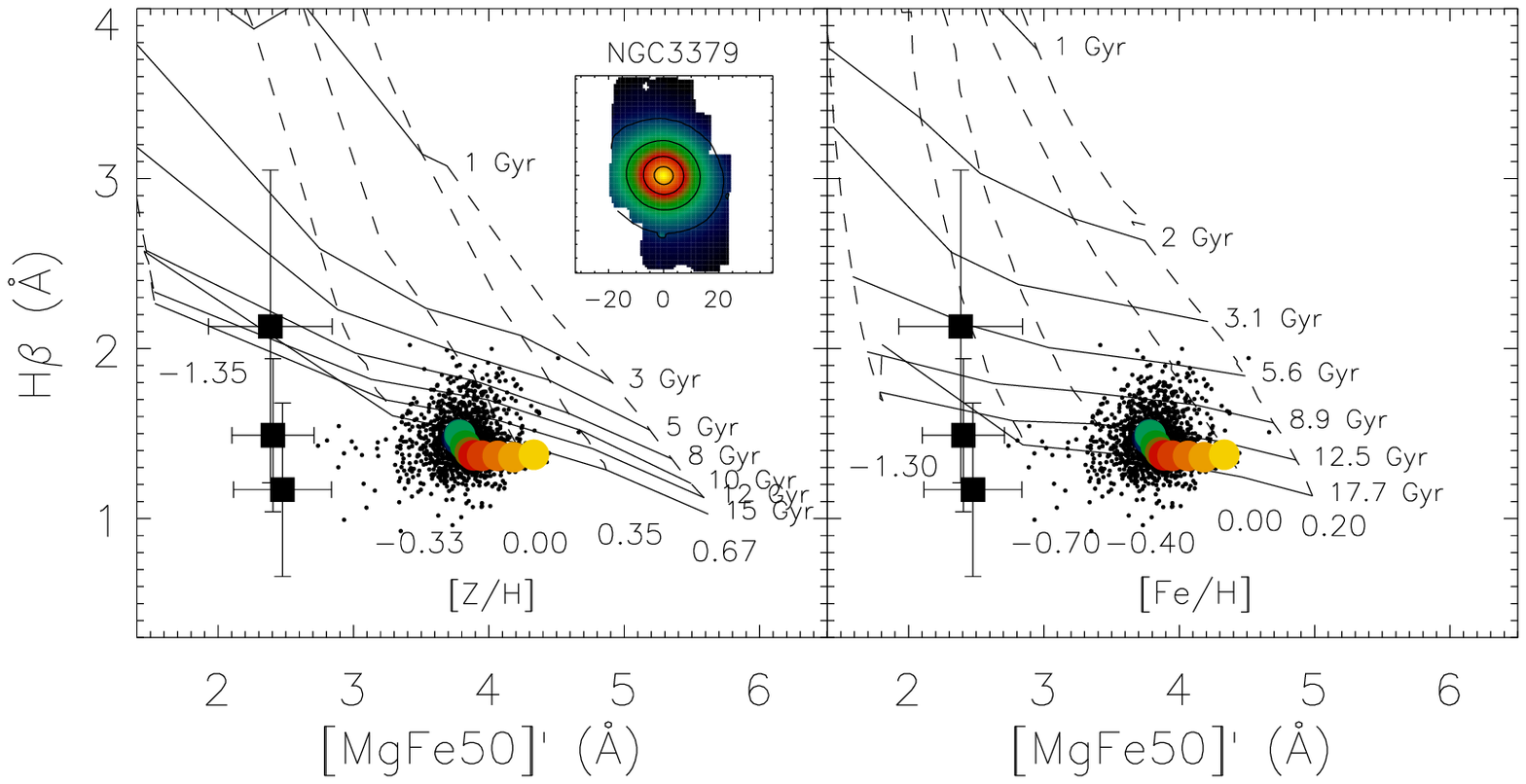,angle=0,width=5.3cm,clip=true}
\end{tabular}
\caption{Left: Line-strength gradients (from top to bottom: H$\beta$,
Fe5015 and Mg\,$b$, in magnitudes) out to 4 $R_e$ in NGC\,3379. Black
triangles denote SAURON data from the original survey \citep{ku06_aw},
and red dots are our new measurements. Right: H$\beta$ index against [MgFe50]$^\prime$, overplotted on the stellar population models of \citet{th03_aw}. Black dots show measurements from the SAURON survey, while the coloured dots are averaged along isophotes (see inset for colour coding). The black filled squares are our measurements at large radii.}
\label{fig:line-strengths}
\end{figure}

We obtained line-strength measurements following the procedures
outlined in \citet{ku06_aw}. We find that the slope of the
line-strength gradients remains constant out to at least 4 $R_e$,
although our values of Fe5015 are not consistent with this trend
(Fig.~\ref{fig:line-strengths}). Plotting our measurements on the
stellar population models of \citet*{th03_aw}, we find that the
stellar halo population is old ($\sim$ 12 Gyr) and metal-poor (below
20\% solar).

\section{Conclusion}

We showed that by using SAURON as a 'photon collector', we can measure
both the stellar velocity profile and absorption line-strengths out to
large radii in early-type galaxies. We presented our measurements of
NGC\,3379 and modeled its dark halo. In our best-fit model, 41\% of
the total mass within 5 $R_e$ is dark. We will present more elaborate
modeling of the dark halo of NGC\,3379 and comparisons with literature
values in a forthcoming paper (Weijmans et al. in preparation), as
well as a comparable dataset and analysis for the elliptical galaxy
NGC\,821.

\acknowledgements It is a pleasure to thank the organisers for a
stimulating and fruitful conference. We gratefully acknowledge Chris
Benn, Eveline van Scherpenzeel, Richard Wilman and the ING staff for
support on La Palma.
 
The SAURON observations were obtained at the William Herschel
Telescope, operated by the Isaac Newton Group in the Spanish
Observatorio del Roque de los Muchachos of the Instituto de
Astrof\'isica de Canarias.

\end{document}